# Anode current saturation of ALD-coated Planacon® MCP-PMTs

**E.V. Antamanova,[a] I.G. Bearden,[b] E.J. Garcia-Solis,[c] A.V. Harton,[c] V.A. Kaplin,[a] T.L. Karavicheva,[d] J.L. Klay,[e] Yu.A. Melikyan,[a, 1] D.V. Serebryakov,[d] M. Slupecki[f] and W.H. Trzaska[f]**

[a] *National Research Nuclear University MEPhI (Moscow Engineering Physics Institute), Kashirskoe shosse 31, Moscow, 115409, Russia*

[b] *Niels Bohr Institute, University of Copenhagen, Nørregade 10, 1017 Copenhagen, Denmark*

[c] *Chicago State University, 9501 S. King Drive, Chicago, IL 60628, United States*

[d] *Institute for Nuclear Research of the Russian Academy of Sciences, V-312, 60-letiya Oktyabrya prospect 7a, Moscow, 117312, Russia*

[e] *California Polytechnic State University, 1 Grand Avenue, San Luis Obispo, CA 93407, United States*

[f] *University of Jyväskylä, P.O. Box 35, FI-40351 Jyväskylä, Finland*

*E-mail*: ymelikyan@yandex.ru

ABSTRACT:

We have measured and compared the characteristics of ALD-coated Planacon® MCP-PMTs (XP85112/A1-Q-L) with their non-ALD counterparts (XP85012/A1-Q). While the later show excellent performance, the ALD-coated sensors have surprisingly low current saturation levels (~two orders of magnitude lower than expected) and extremely high gain recovery time (more than 7 orders of magnitude higher than expected). We suspect that these problems might be caused by the unexpected side-effects of the ALD process. To make a definite conclusion, more samples need to be tested, preferably from different production runs. If our observation were confirmed, it would mean a serious technological setback for ALD-coated MCP-PMTs.

KEYWORDS: MCP-PMT, Atomic Layer Deposition, Anode current saturation

---

[1] Corresponding author

## Contents

## 1. Introduction

Microchannel plate-based photomultiplier tubes (MCP-PMTs) have excellent timing properties and low sensitivity to the magnetic field [1]. At the same time their use in high-energy physics experiments was restricted by a relatively low limit on the integrated anode charge (IAC) of about 0.1 C/cm$^2$ [2, 3]. This restriction comes from the photocathode ageing, caused by the ion backflow [4] and/or by the influence of the neutral molecules from residual gases [5]. To overcome this problem, different technologies have been recently implemented:

- ceramic insulators of the cathode chamber volume [6];
- thin aluminium foil separating two MCPs forming a chevron configuration [6];
- photocathode improvements [7];
- thorough electropolishing and electron scrubbing of MCPs [8];
- atomic layer deposition (ALD) forming resistive, secondary-emission, and electrode layers on the inner pores of MCP [2, 8].

Application of ALD technique to Planacon® photosensors manufactured by Photonis USA Pennsylvania, Inc. was shown to increase IAC up to 10 C/cm$^2$ or more [9]. This achievement was essential to consider ALD-coated MCP-PMTs (in particular, the Planacons) for the use in future high-energy physics experiments where photosensors with high IACs are needed, including the PANDA DIRC detector [10] and the timing system of ATLAS Forward Proton detector [11]. For applications with IAC below 0.1 C/cm$^2$, like the FFD detector of NICA MPD [12], non-ALD MCP-PMTs are preferred.

To evaluate potential photosensors for the Cherenkov subsystem (T0+) of the Fast Interaction Trigger (FIT) detector for the upgrade of the ALICE experiment at CERN LHC [13, 14], we have tested two samples of ALD-coated Planacon® XP85112/A1-Q-L MCP-PMT. Photosensors for T0+ require exceptionally good timing (<20 ps), compact sizes (thickness <27 mm), high geometrical efficiency (>80%), ability to operate in a magnetic field up to 0.5 T, ability to achieve IAC>0.6 C/cm$^2$ with not more than twofold decrease in Q.E. and linearity in a wide dynamic range reaching average anode current (AAC) values up to 250 nA/cm$^2$. The main aim of the test was to measure the dependence of MCP-PMT gain versus AAC. This parameter is relevant for FIT but was not specified by the manufacturer and we did not find it in the literature.

## 2. The experimental set-up and procedure

Planacon® XP85112/A1-Q-L MCP-PMT is a square-shaped multianode device with a chevron stack of two 10 μm pore microchannel plates. The external size of its housing is 59x59 mm$^2$, 23 mm thick. The photocathode is 53x53 mm$^2$. The anode is divided into 64 pixels, each with individual output [15]. The sensors tested by us were custom-modified by combining the 64 anodes into four equal groups so that one device represents four independent photomultiplier channels (quadrants) with a common bias [14]. All quadrants were illuminated by 470 nm LED light in pulsed mode with a constant intensity and variable repetition rate. It is known that, unlike conventional PMTs, anode current saturation limit of MCP-PMT is proportional to the illuminated area [16]. To ensure high illumination uniformity (see figure 1), the light from LED was reflected from a matte diffusion screen. Reference Hamamatsu R11410-20 PMT was used to monitor the intensity of LED light during the test – it was illuminated via a small (<1 cm$^2$) semi-transparent area of the screen. The low-resistance base of this PMT ensured its linear response for AAC up to ~8 μA, while neutral density filter was used to attenuate the light not to reach this limit even during the illumination at high rates.

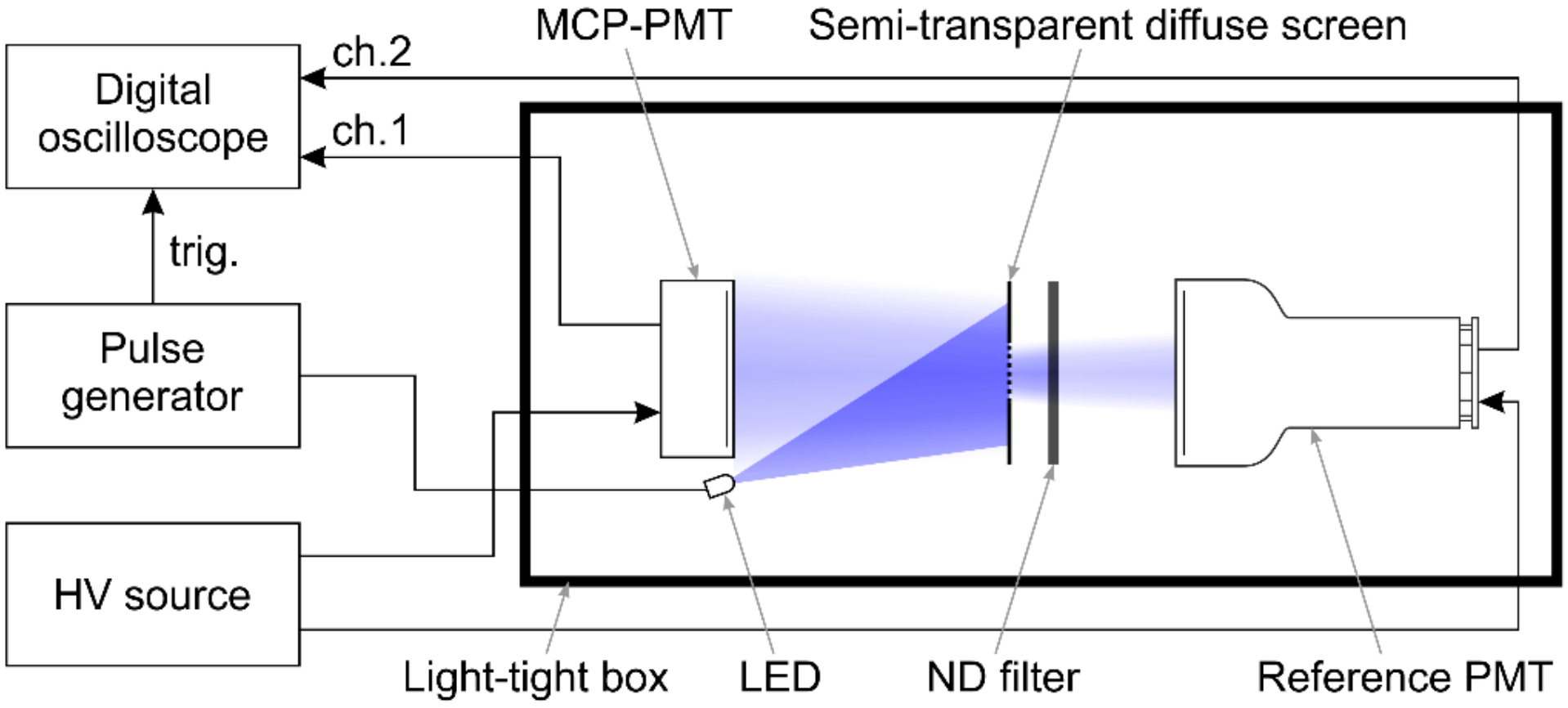

**Figure 1**. The experimental set-up to measure average anode current saturation limit of MCP-PMT.

The tested devices have serial number #9002094 and #9002095 (hereafter #94 and #95) with the MCP resistances 6 MΩ and 8 MΩ correspondingly (values stated in the spec sheets – real values depend to some extent on the MCP current). MCPs of such low resistance values were selected by the manufacturer for the highest anode current saturation limit, which was expected to be equal to 7-10% of the MCP strip (recovery) current [16, 17]. The resistors used in the voltage divider circuit for MCP-PMT #94 were 0.36 MΩ between the photocathode and front of the MCP stack, 9.6 MΩ across the MCP stack, and 0.36 MΩ between the back of the MCP and the anode plane. For MCP-PMT #95 the corresponding values were 0.36 MΩ / 6.56 MΩ / 0.36 MΩ. We have used higher resistivity than the recommended 0.1 MΩ / 1 MΩ / 0.1 MΩ to reduce the power consumption of the device.

The main tests were performed at ~10$^4$ – the gain required by the FIT detector – and repeated at the gain of 10$^5$ conforming with the datasheet requirements. Before each measurement the reference PMT was turned on and stabilized for ~0.5 hour. The tested MCP-PMT were off at that time. Signals from one of the quadrants of the tested MCP-PMT and from the reference PMT were digitized by LeCroy WaveRunner 640 Zi oscilloscope with 10…20 Hz readout rate, depending on the illumination rate. The trends of signal amplitude and charge were measured at illumination rates increasing step-wise from 20 Hz to 400 kHz and back to 20 Hz, as shown in figure 2 (b). Consequently, the AAC value was first growing and then instantly dropping to the initial value. The light intensity was chosen to be ~1.2*$10^4$ photoelectrons per pulse, corresponding to pulses with amplitude ~90 mV and charge ~19 pC at the MCP anode. This value shifted slightly at high illumination rates due either to the LED or to the pulser instability. To compensate for these instabilities, the signal amplitude of the tested MCP-PMT was normalized to the reference PMT signal. The average anode current of the tested PMT was calculated as the product of the signal charge averaged over 1000 events and the corresponding illumination rate.

## 3. Results for the ALD-coated MCP-PMT

Figure 2 (a) shows signal amplitude trend for the tested MCP-PMT (black line), signal charge trend for the reference PMT (red line), and the MCP-PMT amplitude trend normalized by the reference charge from the PMT. Since we were interested here in the change of the device's gain versus the illumination rate, the absolute values of all trends are normalized to their average value in the beginning of the test, right before the first illumination rate increase (events #35000-40000). The correspondence between the relative signal amplitude of the MCP-PMT and the stepwise increase of the illumination rate could be reconstructed with the help of figure 2 (b). During the tests, the HV was supplied and the bias current was monitored using CAEN N1470. The extracted voltage values across the MCP stack are plotted in figure 2 (c).

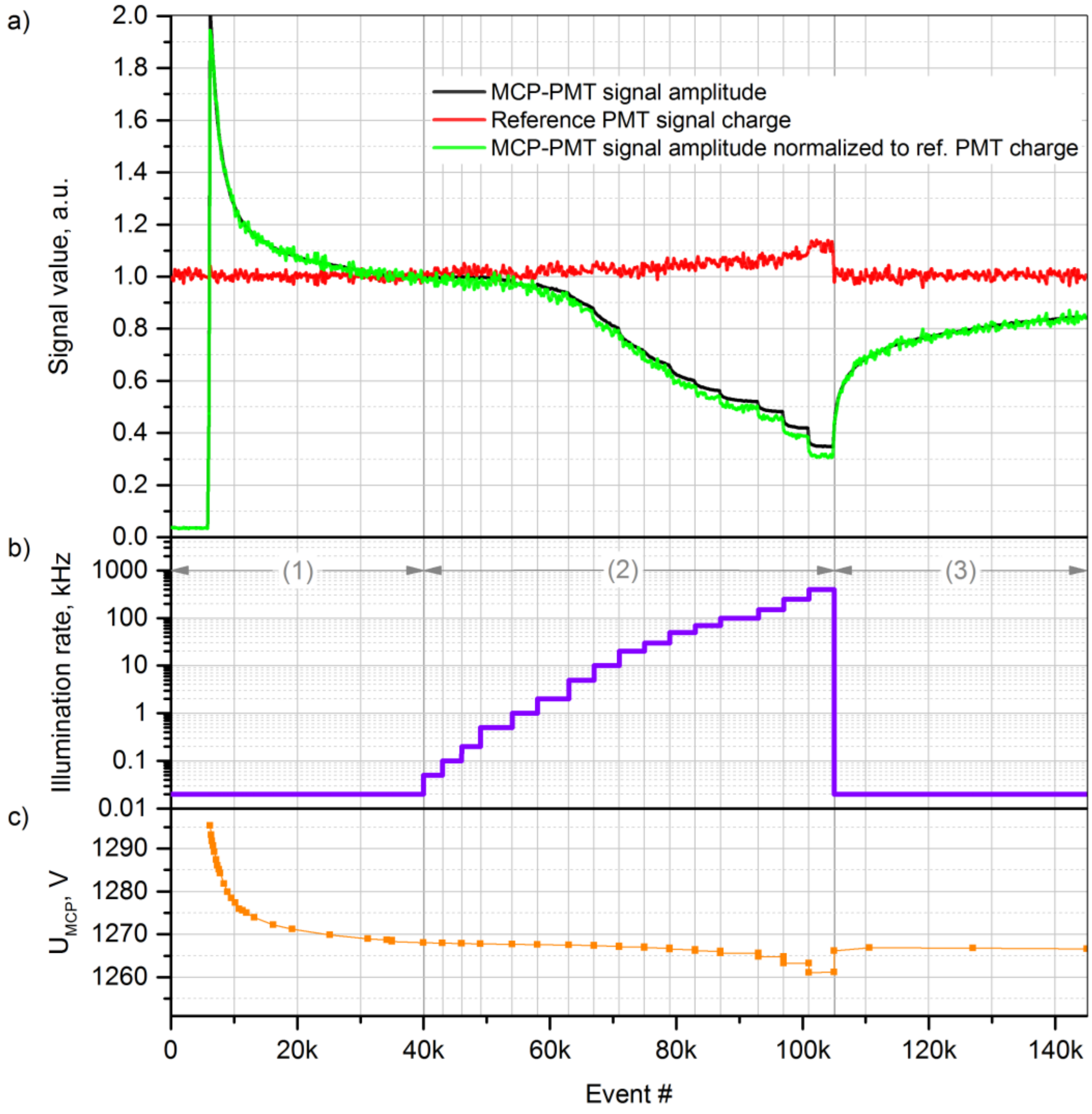

**Figure 2**. a) Relative signal value of the tested MCP-PMT #95 and the reference PMT during the test versus integrated number of counts (digitized pulse waveforms); b) Illumination rate during the test; c) voltage across the MCP stack calculated from the measured bias currents.

The test consisted of three phases, indicated by arrows in figure 2. After the first 11 minutes (6000 events) of the phase 1, the tested MCP-PMT was switched on. It is demonstrated by the sharp rise of the green curve of the signal value on figure 2 (a). The heat-up of the MCP-PMT took the following 64 minutes. During this time its gain was decreasing exponentially due to the negative temperature coefficient of the MCP [15]. It corresponds to a ~27 V decrease in the voltage across the MCP stack.

PMT heat-up phase was followed by phase 2, during which the illumination rate has been increased stepwise up to 400 kHz. Surprisingly, gain decrease became visible already at 1 kHz illumination rate, which corresponds to 2.6 nA/cm$^2$ of AAC. Taking into account that the strip current was equal to 6.6 µA/cm$^2$ in the beginning of phase 1, this level of saturation was two hundred times lower than expected for this device.

Even more surprising was the slow recovery process (marked as phase 3) after the instantaneous drop in the illumination rate from 400 kHz back to 20 Hz. Usually, MCP-PMT recovery takes no more than few milliseconds and is determined by the resistance of the second

MCP in the stack and the capacitance of its saturated region [15]. In the case of MCP-PMT #95 the gain has not fully recovered even at the end of the 67 minutes of the waiting period during the phase 3.

MCP-PMT #94 was tested in the same setup and showed similar characteristics: its AAC saturation became visible at 13 nA/cm$^2$ only and gain recovery took more than 1 hour after its short-term illumination at 400 kHz (350 nA/cm$^2$ or 9.8 μA/PMT).

## 4. Results for the non-ALD MCP-PMT

To ensure that the observed unexpected behaviour of the ALD-coated Planacon® is not caused by any systematic errors or problems with the setup, we have repeated the test using a non-ALD MCP-PMT. In our case it was a Planacon® XP85012/A1-Q MCP-PMT with serial number #9002059 that will be referred to as MCP-PMT #59. #59 had the same modifications as #94 and #95. The only differences were the lack of ALD treatment and a larger pore diameter: 25 μm for #59 instead of 10 μm for #94 and #95. Figure 3 shows a set of curves similar to those presented in figure 2 (a, b), but obtained for MCP-PMT #59.

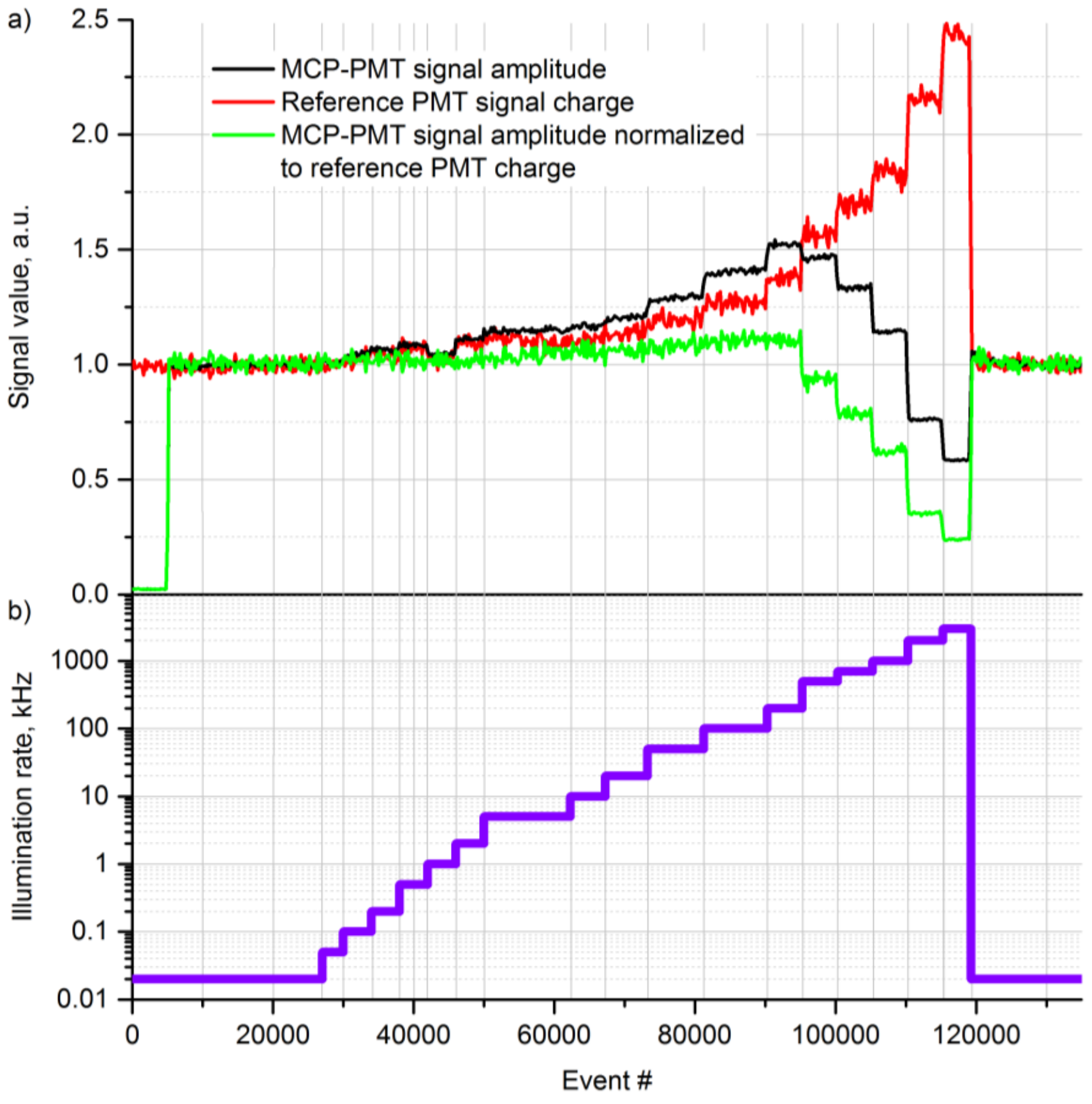

**Figure 3**. a) Relative signal value of the tested MCP-PMT #59, without ALD coating, and for the reference PMT during the test versus the integrated number of counts (digitized pulse waveforms); b) Illumination rate during the test.

According to the spec sheet, the MCP stack of MCP-PMT #59 has 16 MΩ resistance, which is sufficiently high not to require a long heat-up period before the gain stabilization after the ramp up. Consequently, the shape of amplitude trend after turning on the HV supply is flat. As expected, gain decrease became visible only at a high average anode current (730 nA/cm$^2$) followed by an immediate gain recovery after dropping the illumination rate from 3 MHz back to 20 Hz.

## 5. Discussion

Comparison of gain saturation curves of the tested MCP-PMTs is shown in figure 4. Apart from the low values of the saturation level, the experimental curves obtained with the ALD-coated devices have a non-typical shape due to the sloping of the saturated region.

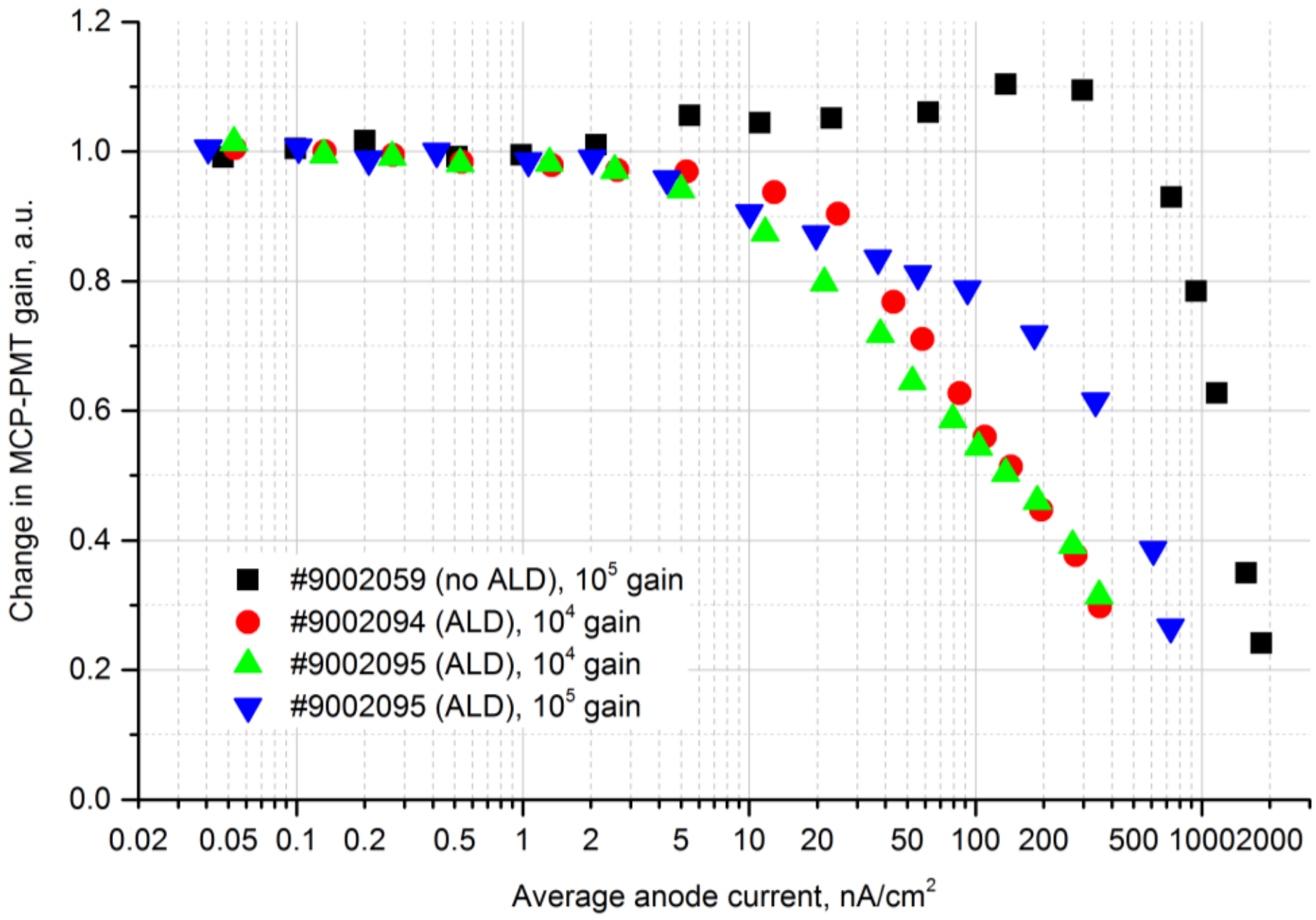

**Figure 4**. MCP-PMT pulse amplitude normalized to the reference PMT charge in (in relative units) versus average anode current in the tested MCP-PMT channel. Total effective area of the MCP-PMT is 28.1 cm2.

It is worth noticing, that the datasheets of the MCP-PMT #94 and #95 state the maximal allowed anode current to be 10 μA/device, or 360 nA/cm$^2$, which has never been increased during the described study. See Table 1 for a summary of the known and measured parameters of the tested devices). To check for possible defects in the ALD devices, another voltage divider base was used, featuring the recommended 0.1 MΩ / 1.0 MΩ / 0.1 MΩ resistivity chain. In addition, the electrical contact between the photocathode and its leads was checked as well. Since none of these tests revealed any reasons for the unexpected behaviour of the ALD-coated MCP-PMTs, we suspect that the low anode current saturation level and the extremely long recovery time after reaching saturation may be related to the side effects of ALD coating. This suspicion is strengthened by the information from our colleagues from the AFP project [11] who observed extended recovery times for an ALD-coated miniPlanacon® device.

Considering that the particular composition and structure of the ALD layers implemented in the Planacon® photosensors is not published, the exact nature of the discussed phenomena remains unclear to us. If the ALD-coating for MCP-PMTs includes a thin $Al_2O_3$ secondary-emission layer [18], it might lead, for example, to a charge build-up inside the bulk of the MCP due to the electron tunnelling effects via the layer. In any case, the observed deficiencies need to be resolved. While it is possible that the low anode saturation level may not be important in low-load applications, the extremely long recovery time of these device could seriously limit their applications.

Table 1. Tested MCP-PMTs parameters (measured or known from the specs sheets)

| Device | #9002094 | #9002095 | #9002059 |
|---|---|---|---|
| Total effective area | 28.1 (5.3x5.3) $cm^2$ | | |
| Pore size | 10 μm | | 25 μm |
| ALD-coating | yes | | no |
| MCP stack resistance according to the datasheet, MΩ | 6 | 8 | 16 |
| Measured MCP stack resistance during the experiment, MΩ | 6.5 | 6.9 | 14 |
| AAC for 10% gain reduction, $nA/cm^2$ | 25 ($10^4$ gain) | 10 ($10^5$ gain) | 800 ($10^5$ gain) |
| Strip current, $\mu/cm^2$ | 12 | 13 | 12 |
| Allowed AAC limit per device, μA | 10 | 10 | 3 |

**6. Conclusions**

We have measured and compared the characteristics of ALD-coated Planacon® MCP-PMTs (XP85112/A1-Q-L) with their non-ALD counterparts (XP85012/A1-Q). While the later show excellent performance, the ALD-coated sensors have surprisingly low current saturation levels (~two orders of magnitude lower than expected) and extremely high gain recovery time (more than 7 orders of magnitude higher than expected). We suspect that these problems might be caused by the unexpected side-effects of the ALD process. To make a definite conclusion, more samples need to be tested, preferably from different production runs. If our observation were confirmed, it would mean a serious technological setback for ALD-coated MCP-PMTs.

As the outcome of this investigation, we have recommended XP85012/A1-Q for the construction of the FIT detector for the upgrade of the ALICE experiment at CERN LHC.